\documentclass[aps,prb,twocolumn,amsmath,amssymb,superscriptaddress,floatfix]{revtex4}
\usepackage{graphicx}
\usepackage{epstopdf}
\usepackage{bm}
\usepackage[usenames]{color}
\bibstyle{apsrev.bib}

\newcommand{\be}{\begin{equation}}
\newcommand{\ee}{\end{equation}}
\newcommand{\beqn}{\begin{eqnarray}}
\newcommand{\eeqn}{\end{eqnarray}}

\begin{document}

\title{Corner contribution to cluster numbers in the Potts model}

\author{Istv\'an A. Kov\'acs}
\email{kovacs.istvan@wigner.mta.hu}
\affiliation{Wigner Research Centre, Institute for Solid State Physics and Optics, H-1525 Budapest, P.O.Box 49, Hungary}
\affiliation{Institute of Theoretical Physics, Szeged University, H-6720 Szeged, Hungary}
\author{Eren Metin El\c{c}i}
\email{elcie@uni.coventry.ac.uk}
\affiliation{Applied Mathematics Research Centre, Coventry University, Coventry, CV1 5FB, United Kingdom}
\affiliation{Institut f\"ur Physik, Johannes Gutenberg-Universit\"at Mainz, Staudinger Weg 7, D-55099 Mainz, Germany}
\author{Martin Weigel}
\email{martin.weigel@coventry.ac.uk}
\affiliation{Applied Mathematics Research Centre, Coventry University, Coventry, CV1 5FB, United Kingdom}
\affiliation{Institut f\"ur Physik, Johannes Gutenberg-Universit\"at Mainz, Staudinger Weg 7, D-55099 Mainz, Germany}
\author{Ferenc Igl\'oi}
\email{igloi.ferenc@wigner.mta.hu}
\affiliation{Wigner Research Centre, Institute for Solid State Physics and Optics, H-1525 Budapest, P.O.Box 49, Hungary}
\affiliation{Institute of Theoretical Physics, Szeged University, H-6720 Szeged, Hungary}
\date{\today}

%\date{\today}

\begin{abstract}

  For the two-dimensional $Q$-state Potts model at criticality, we consider
  Fortuin-Kasteleyn and spin clusters and study the average number $N_\Gamma$ of
  clusters that intersect a given contour $\Gamma$. To leading order, $N_\Gamma$ is
  proportional to the length of the curve. Additionally, however, there occur
  logarithmic contributions related to the corners of $\Gamma$. These are found to be
  universal and their size can be calculated employing techniques from conformal
  field theory. For the Fortuin-Kasteleyn clusters relevant to the thermal phase
  transition we find agreement with these predictions from large-scale numerical
  simulations. For the spin clusters, on the other hand, the cluster numbers are not
  found to be consistent with the values obtained by analytic continuation, as
  conventionally assumed.

\end{abstract}

\maketitle
\section{Introduction}
\label{sec:intr}

The Potts model\cite{potts}, assigning a $Q$-state spin variable to each site of a
lattice with distinct energy contributions between like and unlike spins, describes a
rich set of phase ordering phenomena \cite{wu}. As special cases, it includes the
simpler problem of (bond) percolation for $Q\to 1$ as well as the Ising model
($Q=2$)\cite{baxter}. While these phase transitions are continuous, for sufficiently
large values of $Q$ the transition becomes first order. In two dimensions, this
crossover occurs for $Q> 4$. \cite{wu} While no exact solution is available for the
case of general $Q$, a number of rigorous results regarding the transition
temperatures and critical-point parameters are available from duality and mappings to
vertex models \cite{baxter}. Further results follow from the Coulomb gas mapping
\cite{nienhuis-review}, conformal invariance \cite{cardy_conf} and, more recently,
the framework of Schramm-Loewner evolution (SLE) \cite{sle}.

The relation between the Potts model and percolation \cite{stauffer} as a purely
geometric phase transition becomes apparent from a transformation to a
graph-theoretic problem introduced by Fortuin and Kasteleyn
\cite{Fortuin-Kasteleyn}. There, edges of the lattice are activated with probability
$p$ as in the percolation problem, but each configuration receives an additional
weight proportional to $Q^{N_\mathrm{tot}(F)}$, where $N_\mathrm{tot}(F)$ is the
number of connected components resulting from the bond configuration $F$. For the
choice $p=p_\mathrm{FK}=1-e^{-K}$, where $K$ denotes the reduced coupling, the
percolation transition coincides with the thermal transition of the Potts model and
all magnetic observables can be related to geometric quantities in the percolation
language \cite{coniglio_klein,hu}, such that the fractal structure and connectivity
properties of the Fortuin-Kasteleyn (FK) clusters encode the complete critical
behavior.

One of the basic properties of such percolation configurations is the total number of
clusters. To leading order, this is proportional to the size of the
system. Additionally, however, one expects corrections due to boundary effects
resulting from the presence of surfaces, edges and corners. While, in general, such
correction terms are non-universal, Cardy and Peschel showed that the corner
contribution to the free energy of two-dimensional, conformally invariant systems is
related to the central charge and hence universal \cite{cardy_peschel}. To be
specific, consider a contour $\Gamma$ in the bulk. What is the number of clusters,
$N_{\Gamma}$, which intersect $\Gamma$ when $\Gamma$ is large, but much smaller than
the size of the system? Two of us have recently studied this problem in the
percolation limit $Q\to 1$ and found a logarithmically divergent corner contribution
to $N_{\Gamma}$. \cite{kovacs} Using conformal invariance and the Cardy-Peschel
formula \cite{cardy_peschel} allowed us to calculate this corner contribution
analytically. Full consistency was found with extensive numerical simulations for a
range of different 2D geometries. Here we investigate whether similar results hold
for the more general case of the random-cluster model with arbitrary values of
$Q$. We generalize the analytical calculations and confront the resulting predictions
with large-scale numerical simulations for $Q=1$, $2$, $3$ and $4$, as well as for
the fractional value $Q=0.5$.

Another type of geometrical object in the Potts model are the clusters of like spins
that result from the Fortuin-Kasteleyn construction outlined above with the
alternative choice $p=1$. These also undergo a percolation transition but, in
general, it does not occur at the thermal phase transition point. In two dimensions,
however, both transitions coincide and analogous questions can be studied (such as
fractal dimensions, connectivity properties, etc.) as for the FK clusters. It is
found that, for a given $Q$, FK and spin clusters belong to conformal field theories
of the same central charge \cite{dotsenko_fateev}
\be
c=1-\frac{3(4-g)^2}{2g}
\label{c}
\ee
where $g$ is the Coulomb gas parameter, which is given by the solution of
\be
Q=2+2 \cos(g \pi/2),
\label{Q}
\ee
with $2 \le g \le 4$ for the FK clusters. For the spin clusters one should use the
other solution of Eq.~\eqref{c}, which is given by $g'=16/g$, resulting in $4 \le g'
\le 8$ \cite{deng,janke-geo}. Here, $g'=\kappa$ is just the SLE
parameter\cite{duplantier}. We note that in Eq.~\eqref{Q} the range $4 \le g \le 8$
represents the tricritical branch of the (annealed) site-diluted Potts model
\cite{nbrs}, in which the FK (spin) clusters are expected to be mapped to the spin
(FK) clusters in the critical branch.\cite{stella_vanderzande,duplantier_saleur,qian}
In general, results for the critical FK clusters are conjectured to be related to the
critical spin clusters by analytical continuation, by making the substitution $g \to
g'$. This type of analytical continuation appears to work well on the level of the
fractal dimensions, as shown in a number of numerical investigations
\cite{vanderzande,deng,janke-geo,weigel,dubail1,dubail2}. The universal prefactor in
the area distribution of Ising spin clusters follows the above description as well
\cite{area}.  More recently, however, the three-point connectivities were studied and
the numerical results concerning the spin clusters disagree with the conjecture of
analytical continuation\cite{threepoint}. In the context of corner contributions to
the cluster numbers it is natural, then, to also study the behavior of spin clusters
with $p=1$ or, more generally, the behavior of the continuity of possible cluster
definitions as the bond dilution parameter $p$ is varied.

The rest of the paper is organized as follows. In Sec.~\ref{sec:Potts} we introduce
the bond-diluted model and present the calculation of the corner contribution for FK
clusters in the framework of the random-cluster representation and the arguments of
conformal field theory following the work of Cardy and Peschel. The numerical results are
presented in Sec.~\ref{sec:numerical}. Finally, Sec.~\ref{sec:disc} contains our
conclusions.

\section{Cluster numbers in the Potts model}
\label{sec:Potts}

We consider the $Q$-state Potts model defined by the Hamiltonian \cite{wu}
\be
{\cal H}/k_B T= - K \sum_{\langle i,j\rangle} \delta_{s_i,s_j},
\label{hamilton}
\ee
with the Potts spin variables $s_i=1$, $2$, $\dots$, $Q$, where the summation is over
nearest-neighbor pairs only. We restrict ourselves here to the model on the square
lattice with a total of $n$ sites and $m$ bonds. Following the prescription
introduced by Fortuin and Kasteleyn \cite{Fortuin-Kasteleyn}, the partition function
of the model can be written as
\be Z(Q) \sim \sum_{F} Q^{N_\mathrm{tot}(F)} p^{M(F)}
{(1-p)}^{m-M(F)},
\label{Z}
\ee
where the bond configuration $F$ consists of $N_\mathrm{tot}(F)$ connected components
and has a total of $M(F)$ active edges. Here the bond occupation probability between
neighboring sites with the same Potts state is given by $p=p_\mathrm{FK}=1-e^{-K}$,
such that the clusters of $F$ are FK clusters. In contrast to the Potts model of
Eq.~\eqref{hamilton}, which only makes sense for integer values of $Q$, the
Fortuin-Kasteleyn form \eqref{Z} is valid for arbitrary positive real $Q$. Regular
bond percolation is easily seen to correspond to the limit $Q \to 1$. The mean total
number of FK clusters is
\be
\left\langle N_\mathrm{tot}\right\rangle=Q \dfrac{\partial \ln Z(Q)}{\partial Q}.
\label{N_tot}
\ee
Let us now introduce a contour $\Gamma$ and assume for simplicity that $\Gamma$
runs on top of a sub-set of the bonds. If we fix all spins on $\Gamma$ (in state $1$,
say), but leave the couplings unchanged, the partition function becomes
\be
Z_{\Gamma}(Q) \sim  \sum_{F} Q^{N_\mathrm{tot}(F)-N_\Gamma} p^{M(F)}{(1-p)}^{m-M(F)},
\label{Z_Gamma}
\ee
where $N_{\Gamma}$ is the number of clusters which intersect $\Gamma$. As a result,
\be
\left\langle N_\mathrm{tot} - N_{\Gamma} \right\rangle=Q
\dfrac{\partial \ln Z_{\Gamma}(Q)}{\partial Q}\;.
\label{N_tot-N_Gamma}
\ee
At the critical point, $e^{K_c}=1+\sqrt{Q}$ (see Ref.~\onlinecite{wu}), we can write:
\cite{cardy_peschel,kovacs}
\beqn
\ln Z(Q)&\sim&A f_b(Q), \cr
\ln Z_{\Gamma}(Q)&\sim&A f_b(Q)+L_{\Gamma} f_s(Q) + C_{\Gamma}(Q) \ln L_{\Gamma},
\label{log_Z}
\eeqn
where $A \propto n$ is the total area, $L_{\Gamma}$ is the length of $\Gamma$, and
$f_b$ and $f_s$ are the bulk and surface free-energy densities, respectively. The
latter are non-universal quantities.  The last term in Eq.~\eqref{log_Z} represents
the corner contribution.  Together with Eqs.~\eqref{N_tot} and \eqref{N_tot-N_Gamma}
we hence obtain:
\be
\left\langle N_{\Gamma} \right\rangle=-Qf'_s(Q)L_{\Gamma}+b_{\Gamma}(Q)\ln L_{\Gamma} \;,
\label{N_Gamma}
\ee
with $b_{\Gamma}(Q)=-QC'_{\Gamma}(Q)$. Analogous to the percolation case discussed in
Ref.~\onlinecite{kovacs}, we argue that the partition function $Z_\Gamma(Q)$
decomposes exactly into a product of the partition function $Z^{\rm int}_\Gamma(Q)$
for the interior of $\Gamma$, and $Z^{\rm ext}_\Gamma(Q)$ for the exterior. If there
were only clusters that do not cross $\Gamma$, this property was clearly
fulfilled. Clusters with common points with the boundary, however, are all in the
same Potts state, exactly as for percolation, thus including these does not violate
the product property.  Consequently we can apply the Cardy-Peschel
formula\cite{cardy_peschel} both to the exterior boundary, with corners with interior
angle $\gamma_k$, and to the interior boundary, with $\gamma_k$ replaced by
$2\pi-\gamma_k$.  Using the results of Ref.~\onlinecite{cardy_peschel}, we therefore
deduce that the prefactor of the logarithm in Eq.~\eqref{log_Z} is given by
\beqn
C_{\Gamma}(Q)&=&\dfrac{c(Q)}{24} \sum_k \left[ \left( \dfrac{\pi}{\gamma_k}\right)- \left( \dfrac{\gamma_k}{\pi}\right)\right. \cr
&+&\left. \left(\dfrac{\pi}{2 \pi-\gamma_k}\right)-\left(\dfrac{2 \pi-\gamma_k}{\pi}\right)\right],
\label{cardy_peschel}
\eeqn
where $\gamma_k$ is the interior angle at each corner, and $c(Q)$ is the central
charge as given in Eq.~\eqref{c}.  Using the critical branch of Eq.~\eqref{Q}, we
have:
\be
Qc'(Q)\equiv \beta(Q)=\frac{3}{\pi}\frac{1-16/g^2}{\tan(\pi g/4)}\;,
\label{c'}
\ee
thus
\be
b_{\Gamma}(Q)=\beta(Q) A_{\Gamma}\;,
\label{A}
\ee
where $A_{\Gamma}$ depends on the geometry of $\Gamma$, but does not depend on
$Q$. We summarize the values of $\beta(Q)$ for the cases $Q=0.5$, $1$, $2$, $3$ and
$4$, studied numerically below in the last row of Table \ref{table:1}.

In order to potentially extend these considerations to the case of spin clusters or,
more generally, the case of arbitrary values of the dilution parameter $p$, we consider
the Hamiltonian of the diluted Potts model \cite{murata,coniglio_klein,coniglio}
\be
{\cal H}_\mathrm{dil}/k_B T = {\cal H}/k_B T - J \sum_{\langle i,j\rangle}
(\delta_{\tau_i,\tau_j}-1)\delta_{s_i,s_j},
\label{diluted}
\ee where $\tau_i = 1$, $2$, $\dots$, $P$ is an auxiliary Potts variable, and we take
the limit $P \to 1$. Here, in general, $p=1-e^{-J}$ is different from
$p_\mathrm{FK}$. As an analysis of the renormalization group flows shows
\cite{coniglio_klein}, the critical surface of ${\cal H}_\mathrm{dil}$ is at $K=K_c$
and it contains two fixed points: the FK fixed point at $p=p_\mathrm{FK}$, i.e., at
$J=K_c$, is repulsive and controls the scaling of the critical FK clusters discussed
above. The spin (or Potts) fixed point at $p=p_\mathrm{S} > p_\mathrm{FK}$, on the
other hand, is attractive and controls the scaling behaviour of the spin
clusters. For the purposes of our study, therefore, it is natural to conjecture that
the corner contribution to the spin cluster number is described by
Eqs.~\eqref{N_Gamma}--\eqref{A}, but using the analytical continuation $4\le g'\le 8$
of Eqs.~\eqref{c} and \eqref{Q} corresponding to the value of $Q$ under
consideration. For the Ising model $Q=2$, for instance, we have $c=1/2$ with $g=3$
and $g'=16/3$, thus resulting in $\beta = 7/3\pi$ for the FK clusters and $\beta =
7\sqrt{3}/16\pi$ for the spin clusters, cf.\ the values collected in Tables
\ref{table:1} and \ref{table:2}. In the next section, we shall check these
predictions with numerical simulations.

\section{Numerical results}
\label{sec:numerical}

To test the relations for the corner contribution to the cluster numbers in the
random-cluster model conjectured from conformal field theory, we performed numerical
simulations for a number of different values of $Q$. All simulations were carried out
at the critical point of square-lattice systems of edge length $L$. With the
exception of the line segment at a free boundary discussed below in
Sec.\ref{sec:free}, we employed periodic boundary conditions. For the integer values
$Q = 1$, $2$, $3$ and $4$, our simulations were performed using the Swendsen-Wang
algorithm \cite{swendsen}. For $Q=0.5$ we used a recent implementation of Sweeny's
single-bond method \cite{sweeny} based on a poly-logarithmic connectivity algorithm
\cite{eren_martin13}. For a range of system sizes $64 \le L \le 2048$, we thus
generated at least $10^4$ approximately independent configurational samples each. For
integer $Q$, these spin configurations were subjected to an additional
post-processing step, joining like spins with a probability $p$ to form clusters
including, in particular, the choice $p=1$ corresponding to spin clusters. For
$Q=0.5$, it is not obvious how to construct an analogue of spin configurations, such
that we had to restrict our analysis to the FK clusters there. For each type of
clusters, we then counted crossings with a number of different contours to be
described next.

\subsection{Shapes of the contours}
\label{sec:subsystem}

\begin{figure}[tb]
\begin{center}
\includegraphics[width=3.2in,angle=0]{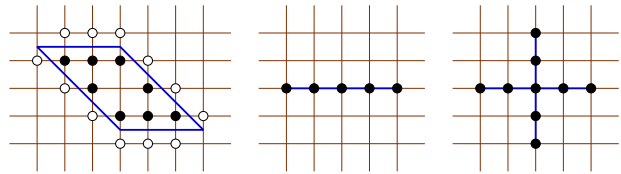}
\end{center}
\vskip -.5cm
\caption{
\label{fig_1} (Color online) Shapes of the contours used in the calculation: sheared squares, line segments in the bulk
and crosses.
}
\end{figure}

With the cluster configurations at hand, we analyzed a number of different contours
$\Gamma$, in particular (sheared) squares, line segments in the bulk and at a free
boundary, as well as crosses.  Three of these shapes are illustrated in
Fig.~\ref{fig_1}. We calculated the corner contributions using the geometric approach
introduced earlier \cite{kovacs_igloi12,kovacs}: for each sample $\left\langle
  N_{\Gamma} \right\rangle$ is calculated in two geometries which have the same
boundary term, but different corner contributions (often it is absent). Hence, the
difference of the two expression provides access to the corner contribution of the
given sample. This approach is useful for cases where strong corrections in the
boundary terms are present.  For the $Q=4$ Potts model studied here, where there are
extra logarithmic corrections \cite{Q4log}, this method spares us to disentangle
these corrections from the logarithmic corner contribution.

%%%%%%%%%% FIG 1  %%%%%%%%%%%%%%%%%%%%%%%%%%%%%%%
%%%%%%%%%% FIG 1  %%%%%%%%%%%%%%%%%%%%%%%%%%%%%%%

\subsubsection{Square and sheared squares}
\label{sec:square}

The first geometry considered here is a square of edge length $L/2$. As shown in
detail in Ref.~\onlinecite{kovacs_igloi12}, the relevant corner contribution can be
computed from comparing two arrangements of subdividing the system into squares or
strips that have the same overall boundary, but the strip configuration has no
corners. Hence, the corner contribution can be found from the difference of the
corresponding cluster numbers. Additionally, one can consider a sheared version of
the square, having an opening angle $\gamma \le \pi/2$ and both its base and its
altitude is given by $L/2$. For this case, the angular dependence for the corner
contribution is found to be:
\be
A_{\Gamma}=\frac{1}{12}\left[4-\pi\left(\frac{1}{\gamma}+\frac{1}{\pi-\gamma}+
    \frac{1}{\pi+\gamma}+\frac{1}{2\pi-\gamma}\right)\right].
\label{sheared}
\ee
For the sheared case, the contour $\Gamma$ cannot run along a sub-set of the
bonds. Instead, we allow it to have an arbitrary position and consider an inner and an
outer layer of spins adjacent to the contour as illustrated in the left panel of
Fig.~\ref{fig_1}. We then consider two types of crossing clusters. Type $(a)$
clusters have common points with sites of both layers, while type $(b)$ clusters are
the $(a)$ clusters and those which have common points with only one of the layers,
however with a weight of $1/2$. This latter case corresponds to $N_\Gamma$ being
averaged over the inner and outer layers. The asymptotic value of the prefactor of
the corner contribution is the same for both type of clusters, but the finite-size
corrections are found to be of different sign, which turns out to be useful for
performing the finite-size extrapolations.

%%%%%%%%%% FIG 2  %%%%%%%%%%%%%%%%%%%%%%%%%%%%%%%
\begin{figure}[tb]
\begin{center}
\includegraphics[width=3.2in,angle=0]{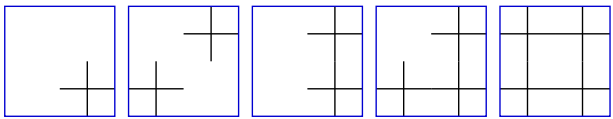}
\end{center}
\vskip -.5cm
\caption{
  \label{fig_2} (Color online) Contours consisting of different numbers of
  crosses. The value of the geometrical factor, $A_{\Gamma}$ in Eq.~(\ref{A}), is
  given for these geometries from left to right by $0$, $0$, $-1/8$, $-1/6$ and
  $-1/4$, respectively, where the values have been normalized to the area of one
  cross.  }
\end{figure}
%%%%%%%%%% FIG 2  %%%%%%%%%%%%%%%%%%%%%%%%%%%%%%%

\subsubsection{Line segments in the bulk}
\label{sec:line}

An even simpler geometry is given by a line segment located in the interior of the
system, which has two exterior angles of each $\gamma=2 \pi$, so that
$A_{\Gamma}=1/8$.  In the geometric approach the line is restricted to lie on top of
a set of lattice points (see the middle panel of Fig.~\ref{fig_1}) and has a length
of either $L/2$ or $L$. In the latter case (with periodic boundary conditions) there
is evidently no corner contribution. In the former case we take two consecutive
segments and calculate $N_{\Gamma}$ for each of them independently. In this way the
clusters which have common points with both segments are calculated twice when we
compare them with the number of clusters for the contour of length $L$.  Thus in this
case the corner contribution for \textit{one} segment of length $L/2$ is just
\textit{half} the number of clusters common to the two segments.

\subsubsection{Crosses}
\label{sec:crosses}

A contour in the form of a cross of edge length $L/2$ is put on a set of lattice
points (see the right panel of Fig.~\ref{fig_1}). In this geometry there are four
exterior angles of size $\gamma=2 \pi$ and another four exterior angles with
$\gamma=\pi/2$, thus the corner contributions cancel out. Hence we also considered
contours $\Gamma$ consisting of two, three or four crosses. In this case, there is a
non-vanishing corner contribution if the number of $2 \pi$ angles differs from the
number of $\pi/2$ angles. The resulting corner contribution for these different
configurations is indicated in the caption of Fig.~\ref{fig_2}. To use the geometric
approach for this setup, we calculate $\left\langle N_{\Gamma} \right\rangle$ in
different geometries, for example comparing the four crosses geometry (rightmost
panel in Fig.~\ref{fig_2}) with the setup of having four independent crosses
(leftmost panel in Fig. \ref{fig_2}). Making use of the possible combinations of the
geometries we obtain several independent estimates of the corner contribution. As a
result of this improved averaging, the crosses setup is usually found to yield the smallest
relative error of the geometries considered here.

\subsubsection{Line segments at a free boundary}
\label{sec:free}

In our last geometry the sample of size $L \times L$ has two free boundaries, whereas
periodic boundary conditions are used in the other direction. The contour is chosen
to be a line segment of length $L/2$ lying at the free boundary. In this case the
magnetization profile is not homogeneous and the Cardy-Peschel formula does not work.
Instead, the logarithmically divergent corner contribution follows from the
properties of the boundary condition changing operator and its prefactor has been
calculated in terms of the Coulomb gas parameter as \cite{cardy01,yu07}
\be
b=\frac{\beta_s}{8}=\frac{1-4/g}{\pi}\sin(\pi g/2).
\label{beta_s}
\ee
%

%%%%%%%%%% TABLE 1  %%%%%%%%%%%%%%%%%%%%%%%%%%%%%%%
\begin{table}[tb]
\caption{Numerical estimates for the prefactor $\beta$
  of the corner contribution for the FK clusters using different contour geometries as compared to the
  conformal predictions of  Eq.~\eqref{c'}.
  The conformal predictions in the last row are $\frac{5\sqrt{3}}{4\pi}$, $\frac{7}{3\pi}$, $\frac{33\sqrt{3}}{25\pi}$
and $\frac{6}{\pi^2}$ for $Q=1$, $2$, $3$ and $4$, respectively.
\label{table:1}}
\begin{tabular}{c|l|l|l|l|l}  %\hline
%\multicolumn{1}{|c|}{}&\multicolumn{3}{|c|}{slab} &\multicolumn{2}{|c|}{pyramid}& \multicolumn{3}{|c|}{wedge} \\ \hline
 $Q$ & 0.5&1 & 2 & 3 & 4 \\ \hline \hline
 squares&$0.589(10)$ & $ 0.689(13) $ & $ 0.742(9) $ & $ 0.734(7) $ & $ 0.669(27) $ \\ \hline
 lines &$0.598(34)$& $ 0.687(12) $ & $ 0.739(11) $ & $ 0.718(11) $ & $ 0.632(4) $ \\ \hline
 crosses &$0.614(30)$& $ 0.692(2) $ & $ 0.742(8) $ & $ 0.730(3) $ & $ 0.650(22) $\\ \hline \hline
  CFT &$0.5933$& $%\frac{5\sqrt{3}}{4\pi}=
  0.6892$ & $%\frac{7}{3\pi}=
  0.7427$ & $%\frac{33\sqrt{3}}{25\pi}=
  0.7278$ & $%\frac{6}{\pi^2}=
  0.6079$
%  & $ 0.6892 $ & $ 0.7427 $ & $ 0.7278 $ & $ 0.6079 $ 
\end{tabular}
\end{table} 
%%%%%%%%%% TABLE 1  %%%%%%%%%%%%%%%%%%%%%%%%%%%%%%%

%%%%%%%%%% FIG 3  %%%%%%%%%%%%%%%%%%%%%%%%%%%%%%%
\begin{figure*}[tb]
\begin{center}
\includegraphics[width=3in,angle=0]{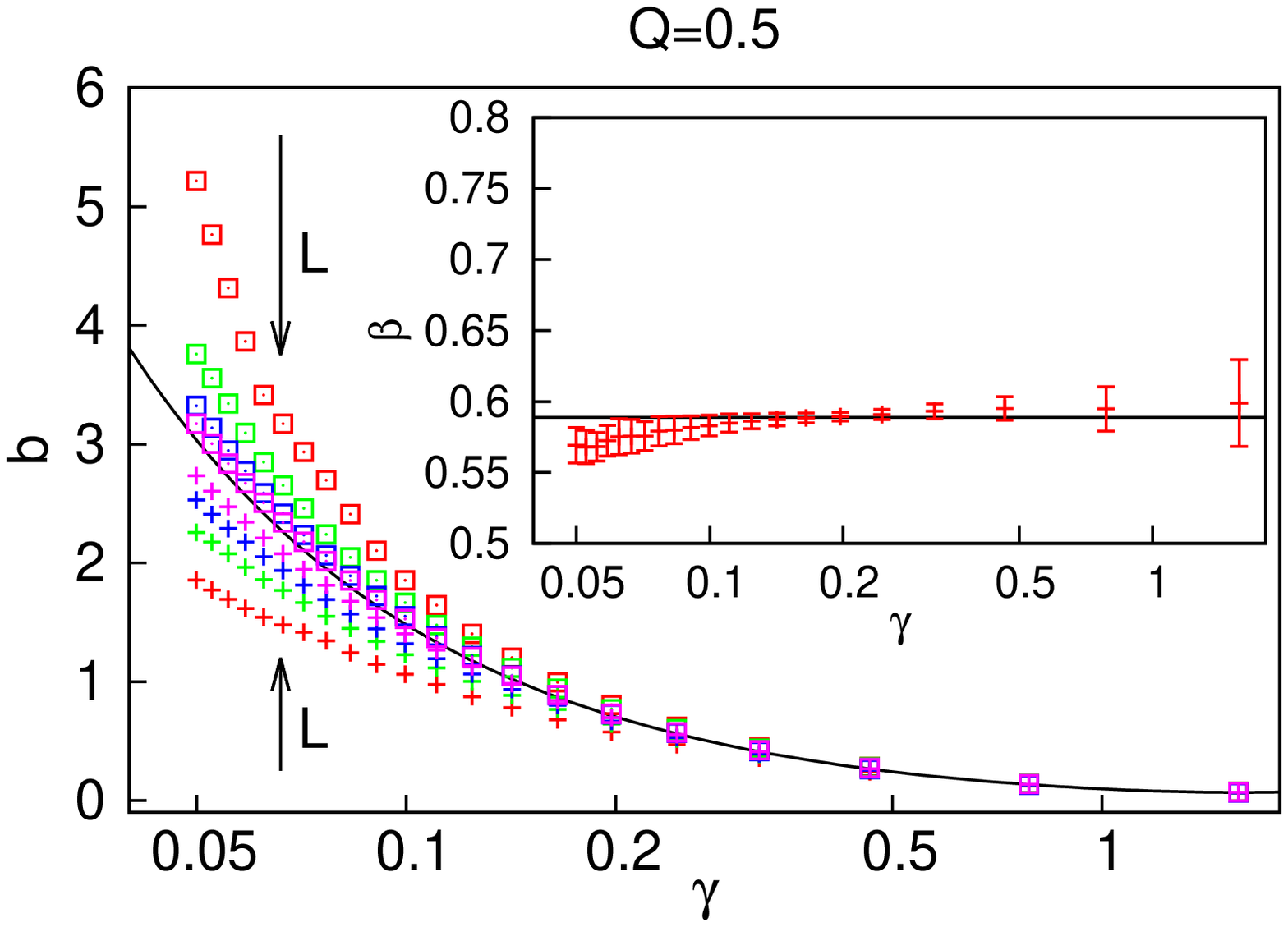}
\includegraphics[width=3in,angle=0]{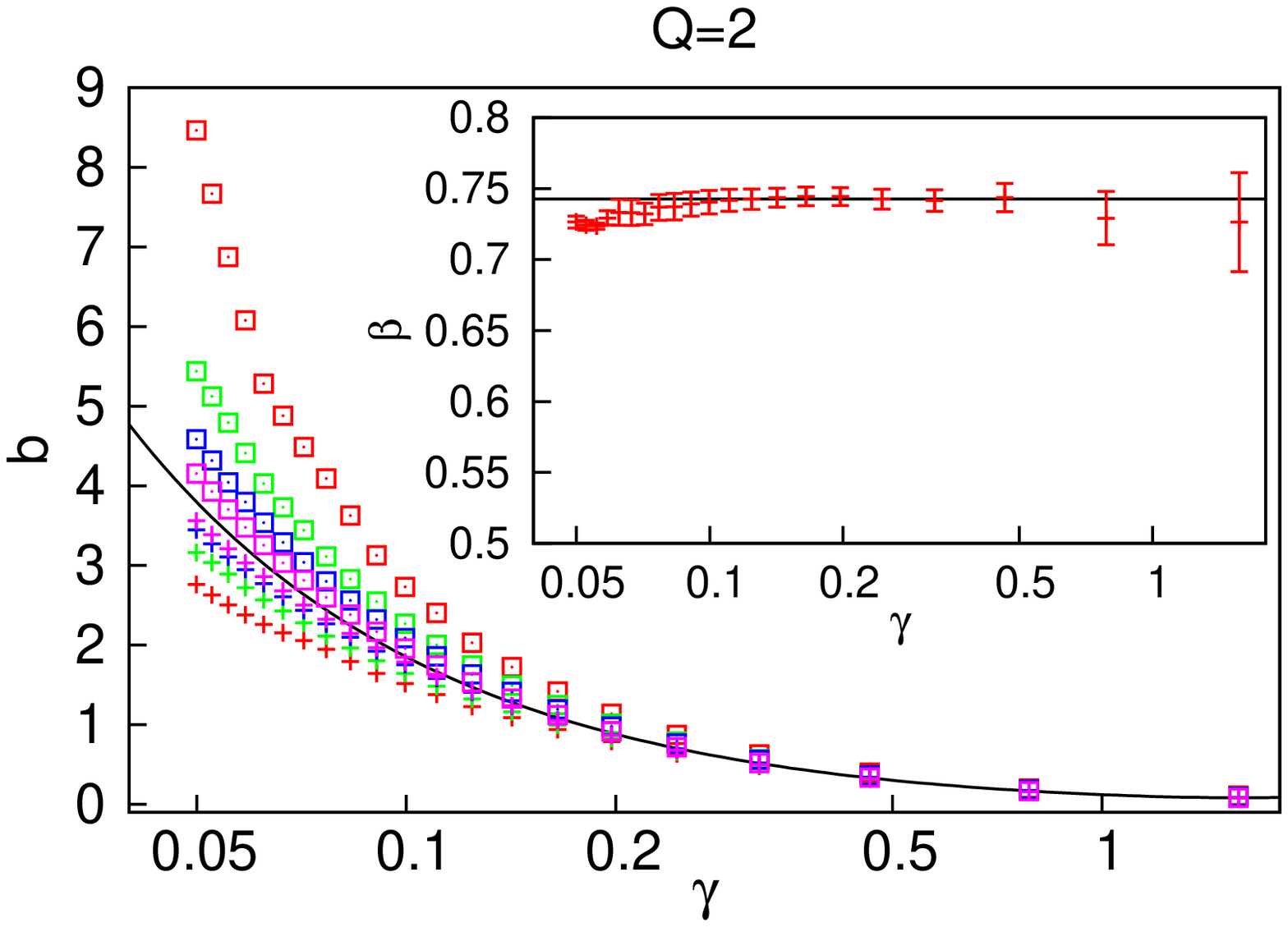}
\includegraphics[width=3in,angle=0]{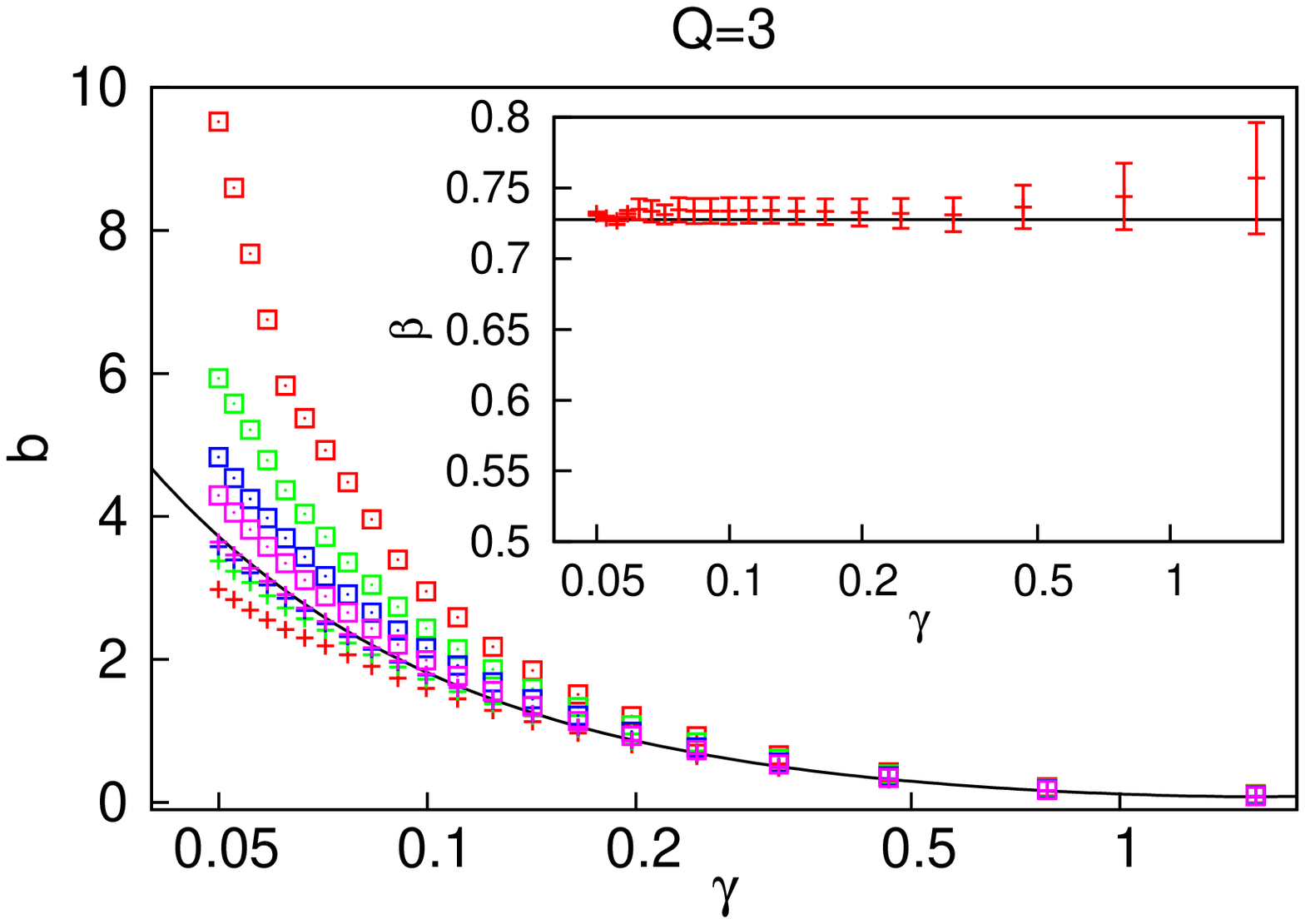}
\includegraphics[width=3in,angle=0]{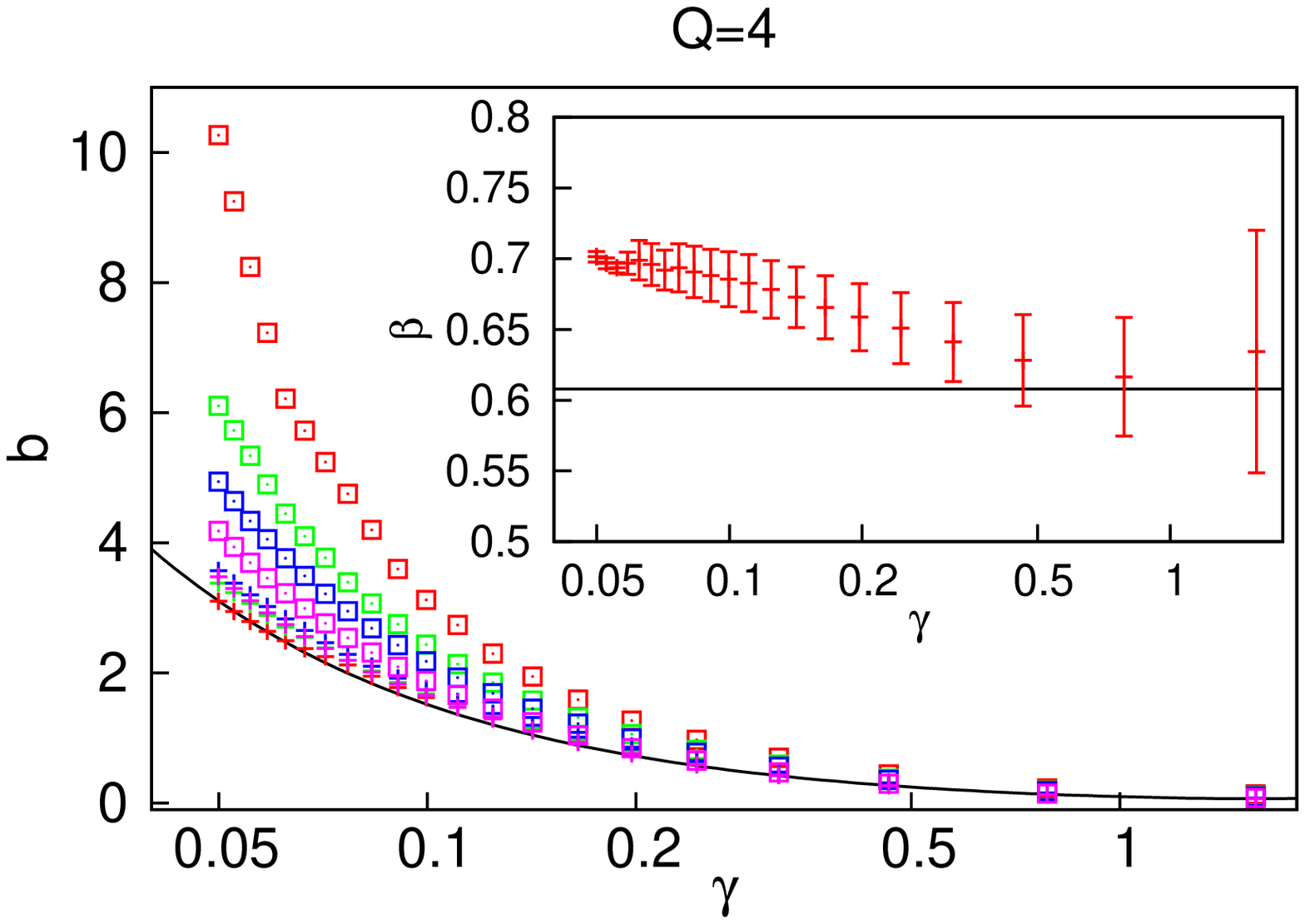}
\end{center}
\vskip -.5cm
\caption{
  \label{fig_3} (Color online) Finite-size estimates ($L=\textcolor{red}{128}$,
  $\textcolor{green}{256}$, $\textcolor{blue}{512}$ and $\textcolor{magenta}{1024}$)
  of the prefactor $b$ according to Eq.~\eqref{A} of the corner contribution to the
  FK cluster numbers with sheared squares as a function of $\gamma$ for type (a)
  ($+$) ($L$ increasing upwards) and type (b) ($\boxdot$) ($L$ increasing
  downwards) clusters. For larger sizes the results are closer to the conformal
  result in Eq.~(\ref{sheared}) that is indicated by the full line.  The insets show the ratio of
  the estimated prefactor $b$ and the angle dependent factor, $A_{\Gamma}$, in
  Eq.~(\ref{sheared}) as a function of $\gamma$.}
\end{figure*}
%%%%%%%%%% FIG 3  %%%%%%%%%%%%%%%%%%%%%%%%%%%%%%%

\subsection{Fortuin-Kasteleyn clusters}
\label{sec:FK}

We first considered the scaling of the corner contributions for the FK clusters which
are directly encoding the critical behavior of the model. For the case of the
(sheared) square, cf.\ Fig.~\ref{fig_1}, for each sample configuration of the Potts
spins resp.\ bonds of the random-cluster representation we averaged over $10^3$
different positions of the contour. For each lattice size up to $L=1024$, we computed
the corner contribution according to the geometric approach as discussed above in
Sec.~\ref{sec:square}. The size dependence of the prefactor $b = b(L)$ was obtained
by logarithmic two-point fits to the data at sizes $L/2$ and $L$, then linearly
extrapolated to the limit $1/L \rightarrow 0$, cf.\ Ref.~\onlinecite{kovacs}.

The results of this analysis for the cases $Q=0.5$, $2$, $3$ and $4$ are shown in
Fig.~\ref{fig_3} for the two used cluster definitions as a function of the opening
angle, $\gamma$. (For $Q=1$ a similar figure can be found in
Ref.~\onlinecite{kovacs}.) In each panel the conformal prediction of
Eq.~(\ref{sheared}) is also indicated by a full line. For not too small values of
$\gamma \gtrsim 0.1$ the finite size corrections are small and the different
estimates for $b(L)$ agree well with the conformal prediction. With decreasing
$\gamma$ the finite-size corrections increase continuously, but for each $\gamma$ the
values extrapolated for $L\to\infty$ are 
in good agreement with the conformal prediction. 
We also estimated the parameter $\beta(Q)$ of Eq.~\eqref{c'} by dividing
the extrapolated prefactor by the angle dependent factor $A_{\Gamma}$ in
Eq.~\eqref{sheared}. The results are shown in the insets of Fig.~\ref{fig_3}.  For $Q \le
3$ the estimates for $\beta(Q)$ are independent of the opening angle $\gamma$ and
within statistical errors the averages are in excellent agreement with the conformal
conjecture. For the limiting case $Q=4$, on the other hand, where the transition is
about to become discontinuous, the agreement is less convincing. This, however, is
not surprising as strong additional logarithmic corrections are expected for this
case \cite{Q4log}. Hence, significantly larger system sizes beyond the reach of
today's computational resources would be required to clearly resolve the asymptotic
behavior.  The extrapolated values of $\beta(Q)$ for squares are summarized in Table
\ref{table:1}.

We repeated the calculation of the corner contribution for the other two types of
contours, namely the line segments (in the bulk) and the crosses, cf.\
Fig.~\ref{fig_1}. For the former, we averaged over all horizontal and vertical
positions of the segments on the lattice. For the crosses, we averaged either over
all positions (for $L=64$) or over $10^4$ random positions (for $L \ge 128$) for each
sample. Using the procedure discussed above, the relation Eq.~\eqref{A} was used to
extract an estimate of $\beta$ from extrapolating $b(L)$ for $1/L\to0$ and dividing
by the angular dependency $A_{\Gamma}$. The corresponding estimates are collected in
Table \ref{table:1}. For $Q < 4$ the estimates for the different contours agree with
each other and all of them are statistically well consistent with the conformal
prediction. For $Q=4$ the estimates for $\beta$ are less satisfactory which, again,
is attributed to the presence of logarithmic corrections.

For the case of the line segment adjacent to a free boundary we again averaged over
all possible positions. Due to the lack of translational invariance, however, these
are by a factor of $L$ fewer than for the bulk case, leading to correspondingly less
precise results. The prefactor, $\beta_s$, is estimated in the same way as for the
bulk segments. The corresponding values are collected in Table \ref{table:3} and
illustrated in the right panel of Fig.~\ref{fig_6}. The conformal conjectures are
shown for comparison.  Again, the numerical and conformal results are in excellent
agreement for all values of $Q$ considered with the exception of $Q=4$ where some
moderate finite-size deviations are seen.

\subsection{Spin clusters}
\label{sec:spin}

%%%%%%%%%% TABLE 2  %%%%%%%%%%%%%%%%%%%%%%%%%%%%%%%
\begin{table}[tb]
\caption{Numerical estimates for the prefactor $\beta$
  of the corner contribution for the spin clusters using different contour geometries as compared to the
  analytic continuation of the conformal predictions of  Eq.~\eqref{c'}.
\label{table:2}}
\begin{tabular}{c|l|l|l}  %\hline
%\multicolumn{1}{|c|}{}&\multicolumn{3}{|c|}{slab} &\multicolumn{2}{|c|}{pyramid}& \multicolumn{3}{|c|}{wedge} \\ \hline
 $Q$ & 2 & 3 & 4 \\ \hline \hline
 squares & $ 0.480(6) $ & $ 0.703(10) $ & $ 0.796(24) $ \\ \hline
 lines & $ 0.486(10) $ & $ 0.659(13) $ & $ 0.653(5) $ \\ \hline
 crosses & $ 0.485(9) $ & $ 0.662(14) $ & $ 0.665(15) $ \\ \hline \hline
 CFT & $\frac{7\sqrt{3}}{16\pi}=0.2412$ & $\frac{11}{12\pi}\sqrt{1+\frac{2\sqrt{5}}{5}}=0.4016$ & $\frac{6}{\pi^2}=0.6079$
%  & $ 0.2412 $ & $ 0.4016 $ & $ 0.6079 $ \\ \hline

\end{tabular}
\end{table}
%%%%%%%%%% TABLE 2  %%%%%%%%%%%%%%%%%%%%%%%%%%%%%%%

As discussed above in Sec.~\ref{sec:Potts}, we also considered corner contributions
to the number of spin clusters crossing a specific contour $\Gamma$. In terms of the
diluted Hamiltonian Eq.~\eqref{diluted} this corresponds to the choice $p=1$ or $J
\to \infty$. As was noted previously \cite{deng,threepoint}, the (attractive) fixed
point corresponding to the behavior of spin clusters is not located at $p = 1$, but
at some $p_S = p_S(Q)$. This is where the smallest scaling corrections are
measured. (For $Q=2$ an unphysical value of $p_S>1$ is observed, but for $Q=3$ and
$4$ one finds $p_S<1$.\cite{deng,threepoint}) In our case, however, we find only
negligible scaling corrections when working directly with the spin clusters, which we
analyze here for system sizes up to $L=2048$. The more general case of arbitrary
$0<p\leq1$ will be discussed below in Sec.~\ref{sec:geometrical}.

\begin{figure}[tb]
\begin{center}
\includegraphics[width=3in,angle=0]{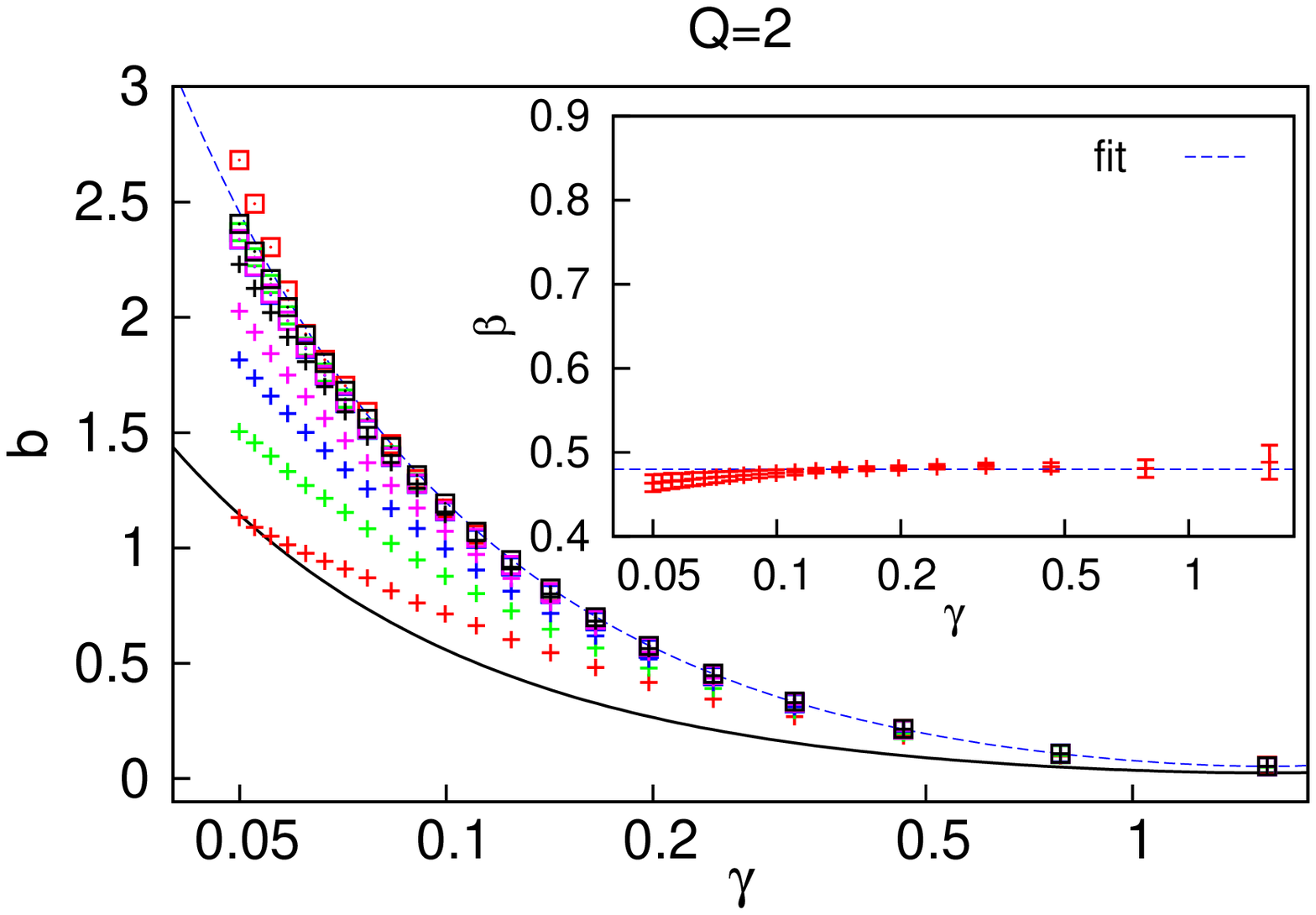}
\includegraphics[width=3in,angle=0]{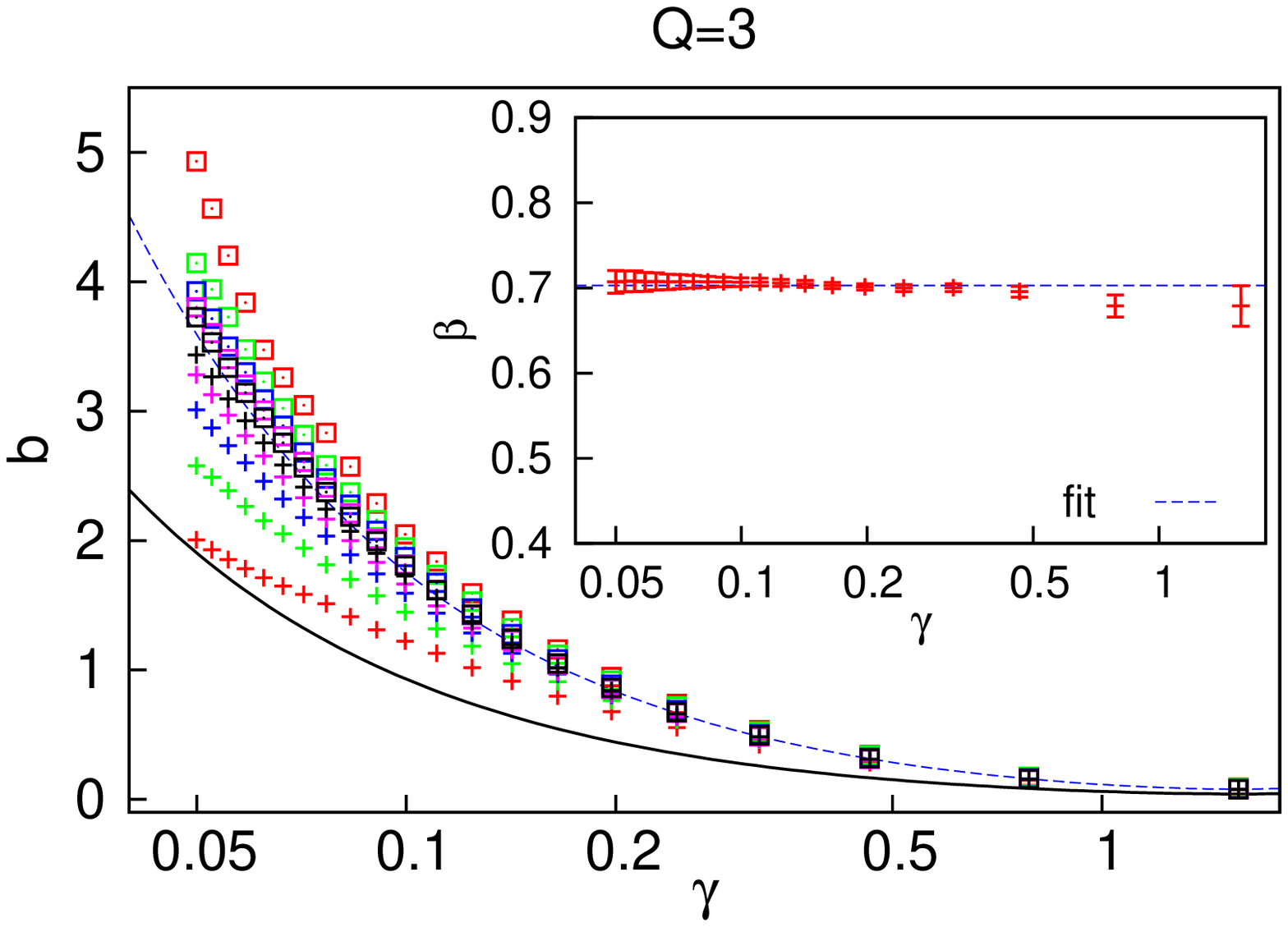}
\includegraphics[width=3in,angle=0]{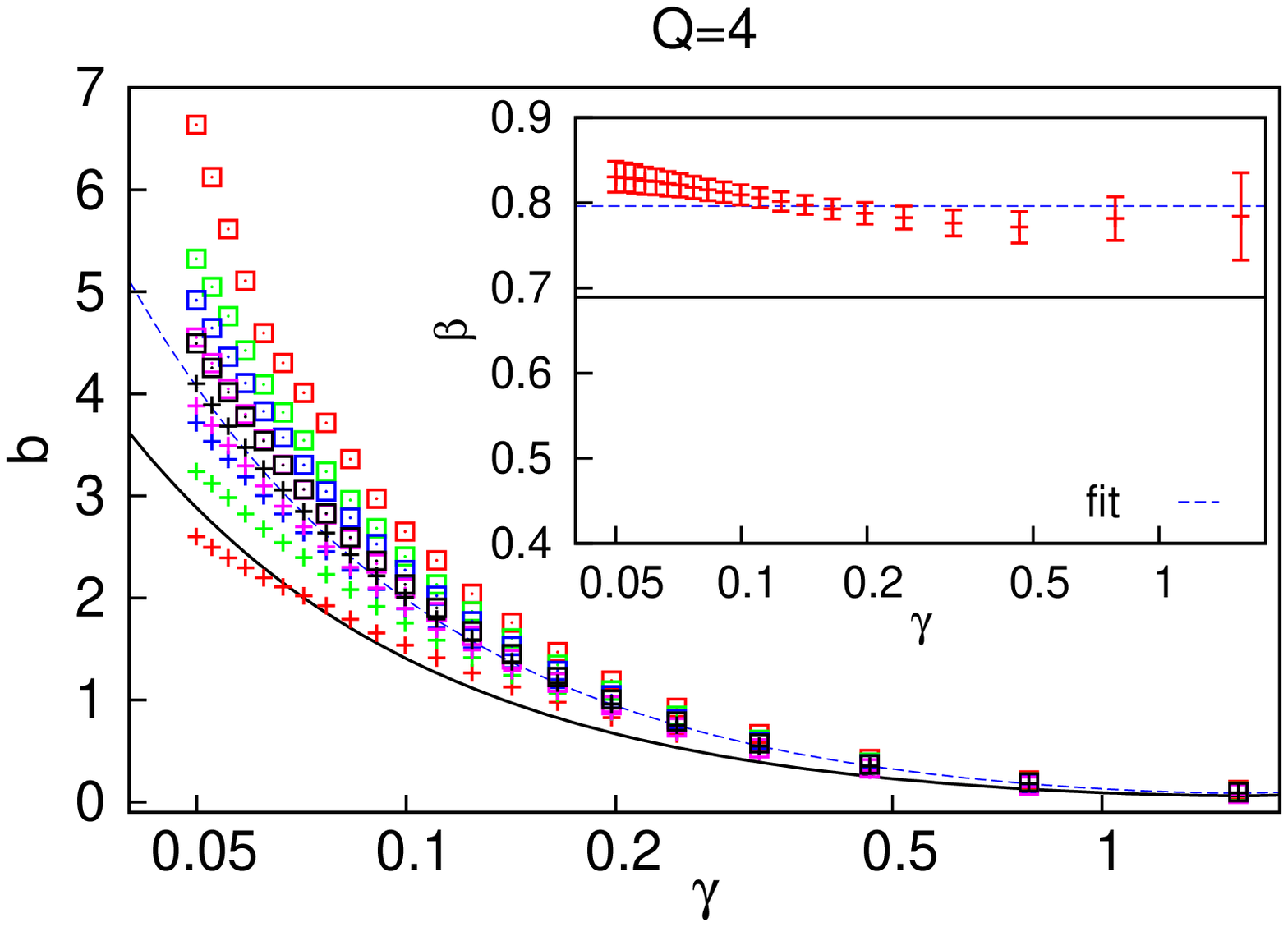}
\end{center}
\vskip -.5cm
\caption{
\label{fig_4} (Color online) 
Numerical estimates for the prefactor $\beta$ of the corner contribution for the spin
clusters for systems up to $2048\times 2048$ spins for sheared squares with opening angle $\gamma$. (The data for $L=2048$
  are presented in black.) The numerically extrapolated prefactors are indicated
  by the dashed lines, while the analytical continuation of the conformal FK results
  are drawn with full lines.  The insets show the ratio of
  the estimated prefactor $b$ and the angle dependent factor, $A_{\Gamma}$, in
  Eq.~(\ref{sheared}).}
\end{figure}

For sheared squares the resulting estimates of the size dependent prefactors for
$Q=2$, $3$ and $4$ are shown in Fig. \ref{fig_4} as a function of the angle $\gamma$.
As for the FK clusters the finite-size corrections for the two types of clusters have
opposite signs, thus allowing for a more efficient extrapolation $L\to\infty$. The
angular dependence of $b(Q)$ indicated by the broken lines in Fig.~\ref{fig_4} is in
agreement with the conformal result of Eq.~\eqref{sheared}, with some deviations
attributed to the presence of logarithmic corrections for the limiting case $Q=4$. As
a result, we can factorize $b(Q)$ as given in Eq.~\eqref{A}. The $Q$ dependent part
$\beta(Q)$, however, does {\em not\/} agree with the result obtained through analytic
continuation of the conformal results for FK clusters, which is indicated by the
solid lines in Fig.~\ref{fig_4}. The numerical estimates for $\beta(Q)$ are shown in
the insets of Fig.~\ref{fig_4} and summarized in Table \ref{table:2}.

%%%%%%%%%% FIG 4  %%%%%%%%%%%%%%%%%%%%%%%%%%%%%%%

%%%%%%%%%% FIG 4  %%%%%%%%%%%%%%%%%%%%%%%%%%%%%%%

For the other geometries, in particular the line segments in the bulk and the
crosses, we arrive at the same observations. The angular dependence is in perfect
agreement with the conformal predictions,  but the $\beta(Q)$ parameters do not agree with
the results of analytical continuation. Notwithstanding these deviations, the
estimates of  $\beta(Q)$ for the three different geometries are in good
agreement with each other, cf.\ the data in Table \ref{table:2}.

Finally, we also considered the corner contributions for spin clusters and a line
segment at a free boundary.  As for the FK clusters we calculated the prefactor
$\beta_s$, and the estimated values are listed in Table \ref{table:3} together with
the conjectured results obtained from Eq.~\eqref{beta_s} by analytical
continuation. Again, the numerical results are different from the conjectured ones,
the numerical data being larger than the conjectured values by roughly a factor of $Q$.

%%%%%%%%%% TABLE 3  %%%%%%%%%%%%%%%%%%%%%%%%%%%%%%%
\begin{table}[tb]
  \caption{Estimates for the prefactor $\beta_s$, calculated for a line segment at a free boundary for $FK$ clusters (upper part)
    and spin clusters (lower part). In both cases the conformal predictions (for spin
    clusters resulting from analytical continuation of the FK results) are presented
    in the second row. 
\label{table:3}}
\begin{tabular}{c|l|l|l|l|l}  %\hline
  $Q$  & $0.5$       & $1$          & $2$            & $3$          & $4$ \\ \hline \hline
  FK   & $1.056(9)$  & $1.101(20)$  & $0.848(29)$   & $0.462(76)$  & $0.13(11)$ \\ \hline
  CFT  & $1.0543$    & $1.1027$     & $0.8488$       & $0.4411$     & \multicolumn{1}{|c}{$0$} \\ \hline  \hline 
  Spin & \multicolumn{1}{|c|}{---}         & \multicolumn{1}{|c|}{$0$}          & $ 1.080(57)$   & $ 1.245(47)$ & $0.54(18) $\\ \hline
  CFT  & \multicolumn{1}{|c|}{---}         & \multicolumn{1}{|c|}{$0$}          & $0.5513$       & $0.4036$     & \multicolumn{1}{|c}{$0$} 
\end{tabular}
\end{table}
%%%%%%%%%% TABLE 3  %%%%%%%%%%%%%%%%%%%%%%%%%%%%%%%

\subsection{Geometrical clusters}
\label{sec:geometrical}

\begin{figure}[tb]
\begin{center}
\includegraphics[width=3.2in,angle=0]{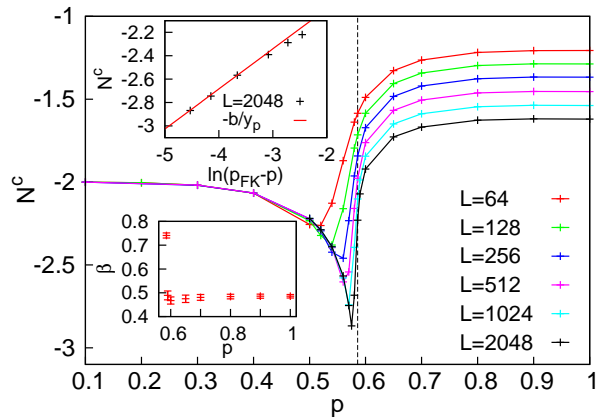}
\end{center}
\vskip -.5cm
\caption{
\label{fig_5} (Color online) Corner contribution $\left\langle N_{\Gamma}^c \right\rangle$ for critical geometrical
$Q=2$ clusters with different values of the bond-dilution probability $p$. The FK
dilution $p_\mathrm{FK}=2-\sqrt{2}$ is indicated by the dashed line.
Upper inset: $\left\langle N_{\Gamma}^c \right\rangle$ vs. $\ln(p_\mathrm{FK}-p)$
for $p<p_\mathrm{FK}$ and for the largest size, $L=2048$. The conformal prediction is shown by a straight line, see Eq.~\eqref{N_p}.
Lower inset: estimated prefactor $\beta$ for $p \ge p_\mathrm{FK}$.
}
\end{figure}

In the previous two sections we have studied the bond-diluted model of
Eq.~\eqref{diluted} for two specific values of $p$ corresponding to the fixed point
of FK clusters, $p=p_\mathrm{FK}$, and to spin clusters at $p=1$.  Generalizing on
this, we might allow for the bond-dilution parameter $p$ to vary between $0 < p \le
1$ and (using the geometric approach) study the corner contribution $\left\langle
  N_{\Gamma}^c \right\rangle$ of the resulting generalized, geometrical clusters at
the critical coupling $K_c$ of the underlying random-cluster model. For the case of
the Ising-like system $Q=2$ and using crosses as contours $\Gamma$ (comparing one and
four crosses in Fig.~\ref{fig_2}, in which case $A_{\Gamma}=-1/4$), we show the
results of such simulations for a range of different system sizes in
Fig.~\ref{fig_5}. Here, for each sample, we averaged over at least $10^3$ positions.

As is clearly seen from Fig.~\ref{fig_5}, for $p<p_\mathrm{FK}$ the finite-size
results converge to a limiting curve, whereas for $p \ge p_\mathrm{FK}$ they grow
with $L$. A closer inspection shows a logarithmic growth, not only at the FK and at the spin cluster
($p=1$) point, but in the complete interval as well. The
prefactor of the logarithm, $b$, is different in (the vicinity of) the FK point,
which was studied above in Sec.~\ref{sec:FK}, and for $p > p_\mathrm{FK}$. Performing
a finite-size analysis in the latter domain, the extrapolated prefactors are found to
be independent of $p$, at least not too close to the FK point, where the crossover
effects are strong, see the lower inset of Fig.~\ref{fig_5}. This observation is in
agreement with the results of the RG analysis that the critical behaviour of
geometrical clusters for $p>p_\mathrm{FK}$ is controlled by the spin cluster fixed
point and hence justifies the use of the spin-cluster limit $p=1$ as a proxy for the
spin fixed point above in Sec.~\ref{sec:spin}.

For the opposite side of the FK point, $p<p_\mathrm{FK}$, the numerical results in
Fig.~\ref{fig_5} indicate that the system is not critical, but has a finite
correlation length, $\xi(p)$, which is divergent at $p=p_\mathrm{FK}$ as $\xi(p) \sim
(p_\mathrm{FK}-p)^{-1/y_p}$. Here the bond-dilution exponent at the FK fixed point,
$y_p$, can be calculated via the Coulomb gas mapping, such that the scaling dimension
$x_p=2-y_p$ is given by\cite{nienhuis-review}
\be
x_p=\frac{1}{8g}(3g-4)(g+4).
\label{x_p}
\ee
For the Ising model with $g=3$ we have $x_p=35/24$ and $y_p=13/24$. In order to
calculate the corner contribution in this non-critical regime, in the second term of
the r.h.s.\ of Eq.~\eqref{N_Gamma} one should replace $L_{\Gamma}$ with $\xi(p)$,
such that we obtain
\be
\left\langle N_{\Gamma}^c \right\rangle=b \ln \xi(p) \simeq \frac{\beta}{4 y_p} \ln(p_{FK}-p)+\mathrm{const}.
\label{N_p}
\ee
In the upper inset of Fig.~\ref{fig_5} we plot $\left\langle N_{\Gamma}^c
\right\rangle$ as a function of $\ln(p_\mathrm{FK}-p)$ and the points are
approximately on a straight line with slope $1.36(2)$, which is compatible with the
theoretical result $\beta/y_p=\frac{56}{13 \pi} = 1.3712$. We have repeated the
calculation for $Q=3$ with similar conclusion, although the error in the slope is
somewhat larger in this case.

%%%%%%%%%% FIG 5  %%%%%%%%%%%%%%%%%%%%%%%%%%%%%%%
%%%%%%%%%% FIG 5  %%%%%%%%%%%%%%%%%%%%%%%%%%%%%%%

%%%%%%%%%% FIG 6  %%%%%%%%%%%%%%%%%%%%%%%%%%%%%%%
\begin{figure*}[tb]
\begin{center}
\includegraphics[width=3in,angle=0]{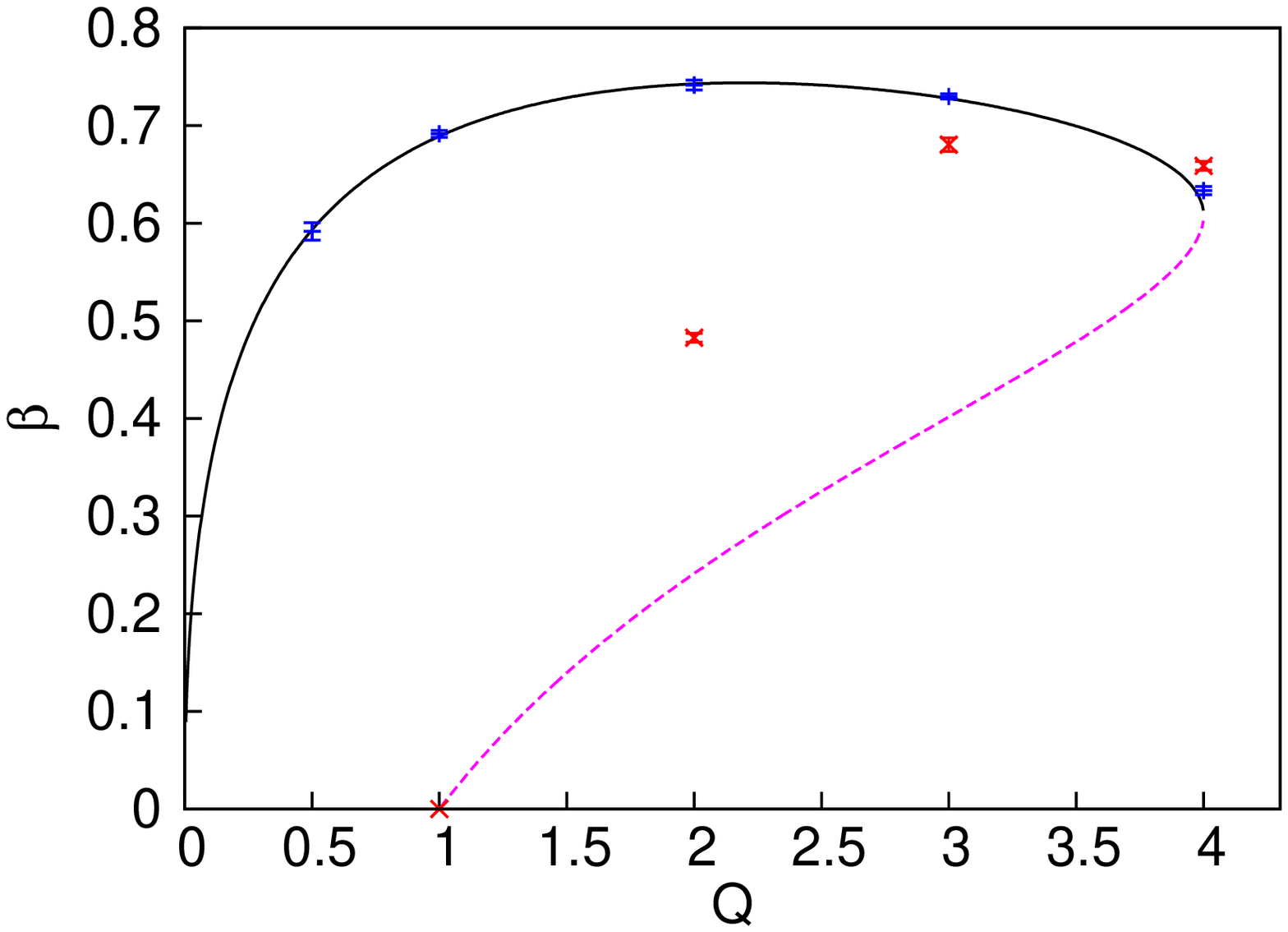}
\includegraphics[width=3in,angle=0]{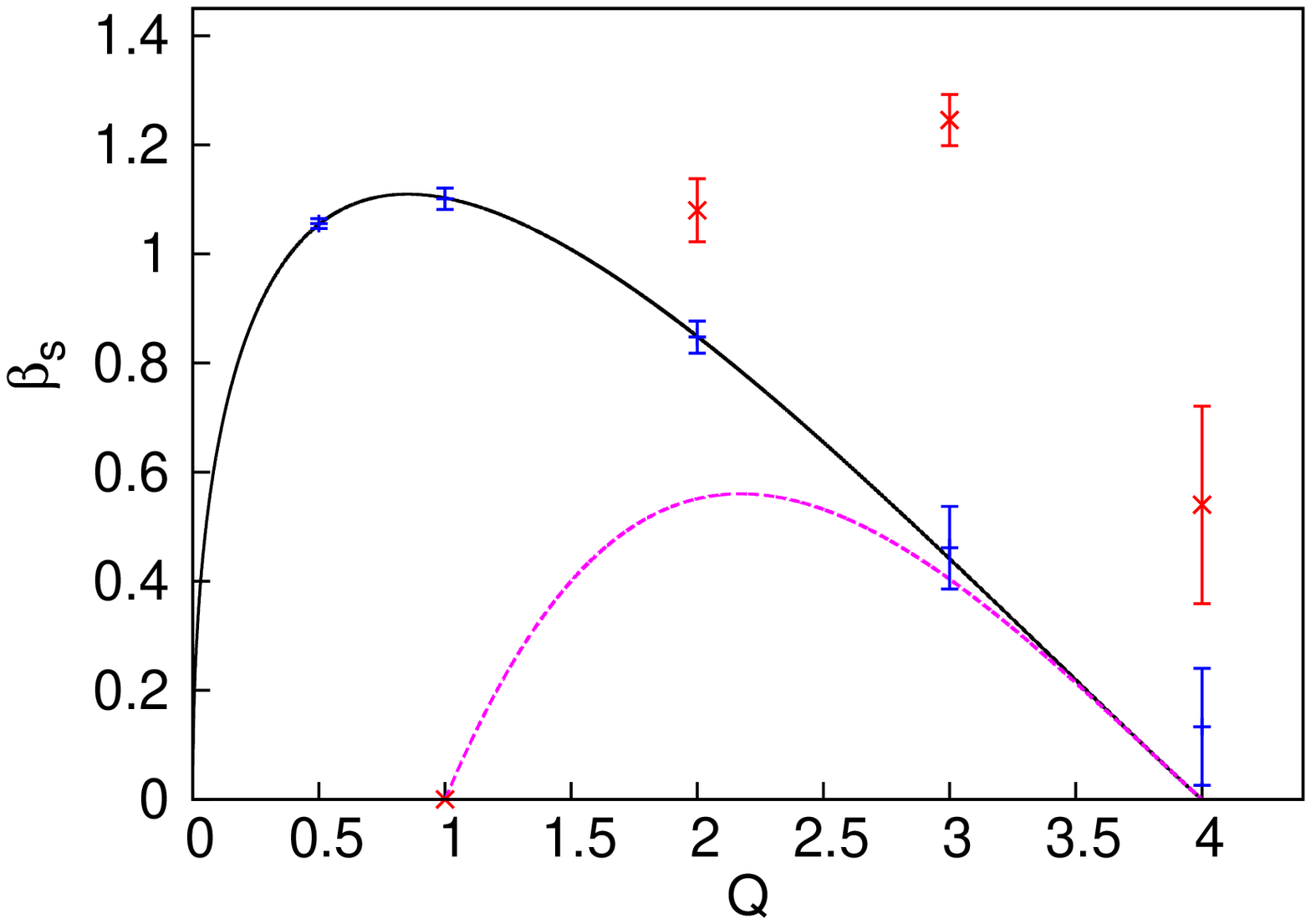}
\end{center}
\vskip -.5cm
\caption{
\label{fig_6} (Color online) Numerical estimates for $\beta$ (left panel) and $\beta_s$ (right panel) for different values
of $Q$ for the FK ($+$) and spin ($\times$) clusters. The full lines represent the conformal conjectures for FK clusters and
the dashed lines are the analytical continuation.}
\end{figure*}
%%%%%%%%%% FIG 6  %%%%%%%%%%%%%%%%%%%%%%%%%%%%%%%

\section{Discussion}
\label{sec:disc}

We have studied the corner contribution of cluster numbers in the $Q$-state Potts
model, both for the critical Fortuin-Kasteleyn clusters and for the critical spin
clusters.  These investigations extend our previous studies at $Q=1$, for
percolation, in which case $\left\langle N_{\Gamma} \right\rangle$ is related to the
entanglement entropy of the dilute quantum Ising
model\cite{lin07,yu07,kovacs_igloi12}. We are not aware of a similar interpretation
for general $Q\ne 1$, although this would be intriguing. For the FK clusters the
corner contribution is expected to be universal and has been calculated via the
Cardy-Peschel formula \cite{cardy_peschel}. Numerical results for different forms of
the contour are in agreement with the conformal conjecture, and the parameter $\beta$
agrees with the conformal results as is illustrated in the left panel of
Fig.~\ref{fig_6}. For spin clusters we follow previous studies finding that the
behavior of critical spin clusters is described by the analytical continuation of the
FK results\cite{vanderzande,deng,janke-geo,weigel} to generalize the conformal
predictions for the amplitudes of the corner contributions. We find, however, that
these conjectures do not agree with the numerical results. Although the angle
dependence of the prefactor follows the Cardy-Peschel formula to high precision, the
parameter $\beta$ differs from the theoretical conjecture as is illustrated in the
left panel of Fig.~\ref{fig_6}. Similar conclusions are obtained when the subsystem
is a line segment at a free surface, in which case the conjectured and measured
values of the parameter $\beta_s$ are given in the right panel of Fig.~\ref{fig_6}.

It is known from previous investigations that the conjecture of analytical
continuation of the FK results to spin clusters works on the level of the fractal
dimensions and the two-point functions
\cite{vanderzande,deng,janke-geo,weigel,dubail1,dubail2,threepoint}. Similarly, the
area distribution of Ising spin clusters follows this conjecture \cite{area}. When,
however, the fine structure of the conformal field theories describing critical
clusters is concerned, such as for the three-point connectivities \cite{threepoint},
the method of analytical continuation breaks down. The present results show that the
universal corner contribution to critical cluster numbers also belongs to this latter
class of properties. As described in Ref.~\onlinecite{kovacs} for the problem of
percolation the universal parameter $\beta$ enters into the expression of a ``corner
probability'' measuring the number of clusters occupying three quadrants of a square,
but have no sites in the fourth one. Also the number of clusters which have common
sites with both halves of a complete line grows logarithmically with $L$, with a
prefactor which is proportional to $\beta$.  These phenomena are shown to be outside
the range of validity of the simple analytical continuation conjecture.

Future directions of research extending the present study include the (annealed)
site-diluted Potts model, that is assumed to feature a tricritical point in the same
universality class as the tricritical point of the diluted Potts Hamiltonian
\eqref{diluted} discussed above \cite{nbrs,janke-geo,qian}. In the presence of
quenched impurities, it is well known that the first-order transition of the model
for $Q>4$ is softened to second order \cite{aizenman_wehr}. For this case,
measurement of $\beta(Q)$ would give access to the central charge of the model which
is of interest as conformal field theories for systems with quenched disorder are
poorly understood. Finally, our investigation could be repeated for models, in which
loops are defined through contour lines of surface growth models, such as the $O(N)$
model.

\begin{acknowledgments}
  This work has been supported by the Hungarian National Research Fund under grant No
  OTKA K75324 and K77629. The research of IAK was supported by the European Union and
  the State of Hungary, co-financed by the European Social Fund in the framework of
  T\'AMOP 4.2.4. A/2-11-1-2012-0001 'National Excellence Program'.  MW acknowledges
  funding by the DFG under contract No. WE4425/1-1 (Emmy Noether Program).
\end{acknowledgments}

\end{document}